%% file: main_PRL_v3.tex
\documentclass[%
twocolumn,
amsmath,amssymb,
aps,
physrev, floatfix
]{revtex4-2}

\usepackage{graphicx}
\usepackage{dcolumn}
\usepackage{bm}
\usepackage{xcolor}
\usepackage{soul}
\usepackage{subfiles}
\newcommand{\cref}[1]{Eq.~\eqref{#1}}
\makeatletter
\let\origaddcontentsline\addcontentsline
\makeatother

\begin{document}
\makeatletter\let\addcontentsline\@gobblethree\makeatother
\preprint{APS/123-QED}

\title{The Inverse Mean First-Passage Time Problem}

\author{Tommer D. Keidar}
\affiliation{%
School of Chemistry, The Center for Computational Molecular and Materials Science, The Center for Physics and Chemistry of Living Systems, Tel Aviv University, Tel Aviv, Israel
}%

\author{Shlomi Reuveni}%
 \email{Contact author: shlomire@tauex.tau.ac.il}
\affiliation{%
School of Chemistry, The Center for Computational Molecular and Materials Science, The Center for Physics and Chemistry of Living Systems, Tel Aviv University, Tel Aviv, Israel
}%

\author{Igor M. Sokolov}
\email{Contact author: sokolov@physik.hu-berlin.de}
\affiliation{
 Institut f\"ur Physik, Humboldt-Universit\"at zu Berlin, Newtonstra{\ss}e 15, D-12489 Berlin, Germany
}%

\date{\today}

\begin{abstract}
We consider the mean first-passage time $\tau(x)$ for an overdamped particle in a one-dimensional potential landscape, and address the problem inverse to the textbook one: Given $\tau(x)$ as a function of starting distance, we look for a potential that realizes it. We derive the conditions on $\tau(x)$ under which such a potential exists. We show that while the power-law dependences $\tau(x)\sim x^\alpha$ with $\alpha \leq 2$ may be realized by monotonic potentials, any realizable $\tau(x)$ with superquadratic growth requires potential wells arbitrarily far from the target; $\alpha = 2$ separates sliding from trapping-dominated motion regimes. 
\end{abstract}

\maketitle

Since Smoluchowski formulated reaction-rate theory \cite{Smoluchowski1918}, the mean first-passage time (MFPT) of an overdamped diffusing particle in a conservative force field has been a central quantity in statistical physics \cite{Farkas1927, Kramers1940, pontryagin1933, Hanggi1990}. More broadly, first-passage observables \cite{Redner_2001, Bray2013, Metzler2014, grebenkov2024} govern the timing of events in systems ranging from chemical reactions \cite{scher2021unified,Benichou2014} and molecular search \cite{berg1981, Kolomeisky2011} to neuronal signaling \cite{gerstein1964}, population dynamics \cite{Ovaskainen2010}, and finance \cite{Chicheportiche2014, Majumdar2008}. In the usual forward problem, one asks how a given potential landscape controls the resulting first-passage dynamics, including the dependence of the MFPT on the initial position.

\begin{figure}[t]
\centering
\includegraphics[width=0.485\textwidth]{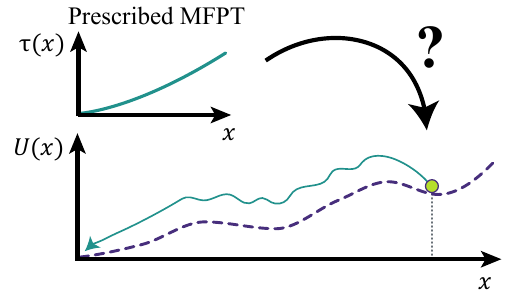}
  \caption{\textbf{The inverse mean first-passage time problem.} Given a prescribed function $\tau(x)$, construct a potential that realizes it as the mean first-passage time of a one-dimensional overdamped diffusion process.}
  \label{fig:Intro}
\end{figure}

Simple examples already suggest a rich relation between landscape shape and first-passage behavior. Denoting by $\tau(x)$ the mean first-passage time of an overdamped Brownian particle to reach the origin from an initial position $x$, a linear potential, corresponding to a constant drift toward the target, gives $\tau(x)\sim x$ \cite{Redner_2001}, whereas a logarithmic potential yields $\tau(x)\sim x^2$ \cite{Bray2000}. This naturally raises an inverse question: by shaping the potential, can one realize an arbitrary growth law of $\tau(x)$, and in particular any power law $\tau(x)\sim x^\alpha$ (see Fig. \ref{fig:Intro})? Despite the fundamental nature of this question, general conditions for realizability have remained largely unexplored. 

Related inverse and design problems have been studied in settings distinct from ours. For example, seeking a moving absorbing boundary that reproduces a prescribed first-passage-time distribution from a single starting point was discussed in~\cite{Zucca2009,chen2011,ettinger2014,abundo2025}. Another line of work optimizes potential landscapes to reduce the MFPT, revealing counterintuitive effects such as a finite barrier shortening, rather than lengthening, the passage time~\cite{Palyulin_2012,Chupeau2020}. The question posed in the present work shifts the focus from reproducing a temporal distribution, or optimizing a single passage time, to finding the potential landscape giving rise to a prescribed function $\tau(x)$ for the MFPT profile. Building such potential landscapes and measuring MFPT on them can be done experimentally, using feedback traps \cite{Cohen2005, Gavrilov2017} or holographic optical tweezers \cite{Grier2006}.

In this Letter, we solve the inverse problem for MFPT profiles in conservative potential landscapes $U(x)$ and uncover a sharp constraint on their asymptotic growth. Note that in what follows, the word ``profile" will always refer to $\tau(x)$ while the word ``landscape" will be used for describing the $U(x)$ dependence.

We first show that the landscape realizing a prescribed $\tau(x)$-profile is unique, up to an additive constant, and is given in closed form, provided the following admissibility condition $\int^\infty dx/\tau'(x)=\infty$ holds. We then show that this criterion forbids $\tau(x)$ from growing faster than quadratically whenever the landscape is monotone beyond some distance. In particular, no monotone landscape can produce a strict superquadratic power law $\tau(x)= Ax^\alpha$ with $\alpha>2$. Quadratic scaling thus emerges as a sharp ceiling for slide-like transport. 

We further show that superquadratic behavior can still be realized as an apparent power law on logarithmic scales, and we explicitly construct the corresponding potentials. Remarkably, these landscapes differ qualitatively from those producing $\alpha\leq 2$: superquadratic growth requires local wells that transiently trap the particle. Our results, therefore, reveal a fundamental transition from sliding to trap-controlled transport.

\textit{The inverse mean first-passage time problem.---}We start by considering a particle, whose location $x$ obeys a 1-dimensional overdamped Langevin equation
\begin{equation}
    \dot{x}=\mu F(x)+\sqrt{2D}\,\xi(t),
    \label{eq:Langevin}
\end{equation}
with $F(x)$ being the force acting on the particle, $\mu$ its mobility, and $D$ the diffusion coefficient. The noise $\xi(t)$ is a white, Gaussian noise, with $\langle\xi(t)\rangle=0$ and $\langle\xi(t)\xi(t')\rangle=\delta(t-t')$. The force obeys $F(x)=-dU/dx$, with $U(x)$ being a potential. We will also assume that the force is continuous and does not diverge for any $0<x<\infty$. The diffusion coefficient obeys the Einstein relation $D=\mu k_BT$ with $k_B$ being Boltzmann's constant and $T$ the bath temperature. Throughout this Letter, we will use a system of units such that $\mu,\,k_BT=1$, in which the potential is measured in units of $k_BT$.

We will start by reviewing known results for the forward problem. We will consider the case where the particle starts its motion at some $x\geq 0$, in the presence of an absorbing boundary at $x=0$. The MFPT to the boundary obeys the backward Kolmogorov equation \cite{pontryagin1933, Redner_2001}
\begin{equation}\label{eq:back_Kolmogorov}
    \frac{d^2\tau}{dx^2}-\frac{dU(x)}{dx}\frac{d\tau}{dx}=-1.
\end{equation}
The solution to this equation is given by
\begin{equation}\label{eq:integral_MFPT}
    \tau(x)=\int_0^x dy\,e^{U(y)}\int_y^\infty dz\,e^{-U(z)}.
\end{equation}
The boundary conditions leading to \cref{eq:integral_MFPT} are
\begin{equation}\label{eq:BCs}
    \tau(0)=0,
    \qquad
    \lim_{x \to \infty} e^{-U(x)}\tau'(x)=0 ,
\end{equation}
namely, an absorbing boundary condition at zero, and a non-trivial boundary condition at infinity. This boundary condition can be derived using the flux-over-population approach, as we review, for completeness, in the Supplemental Material \cite{SM}.

Using \cref{eq:integral_MFPT}, we can find the MFPT in a given potential. For example, for a linear potential $U(x)=Vx$ we get $\tau(x)=(1/V)\int_0^x dy=x/V$. For a logarithmic potential $U(x)=U_0\ln x$ with $U_0>1$, we get $\tau(x)=\left(\int_0^x y\,dy\right)/(U_0-1)=x^2/[2(U_0-1)]$, and for $U_0\leq1$ the MFPT diverges. These two elementary examples show that the large-distance growth of the MFPT is strongly shaped by the confining tail of the potential: a linear potential gives linear growth, while a logarithmic potential gives quadratic growth. This naturally leads to the inverse question: can one design a potential that produces still faster growth, for instance, a cubic law $\tau(x)\sim x^3$?

To try to answer this question, we look at the inverse problem: find $U(x)$ that will generate a prescribed $\tau(x)$. Interpreting \cref{eq:back_Kolmogorov} as an equation for $U(x)$ one gets
\begin{equation}
    U'(x)=\frac{1+\tau''(x)}{\tau'(x)}.
    \label{eq:PotTau}
\end{equation} 
Two corollaries follow. First, $U'(x)$ is fixed pointwise by $\tau(x)$, so the landscape realizing a given MFPT profile is \emph{unique} up to an irrelevant additive constant. Second, observe that if $x'>x$, one must first get from $x'$ to $x$ en route to the target at the origin. Thus, the MFPT is a monotonically increasing function of $x$, $\tau'(x)>0$, and the sign of $U'(x)$ is the sign of $1+\tau''(x)$. Thus, the landscape has a local minimum exactly where $\tau''(x)$ crosses $-1$ from below. Integrating \cref{eq:PotTau} gives the landscape in closed form,
\begin{equation}\label{eq:Uquadrature}
    U(x)=\ln \tau'(x)+\int^x \frac{dy}{\tau'(y)} + C.
\end{equation}

Applying \cref{eq:Uquadrature} to $\tau(x)=A x^\alpha$ with $0< \alpha < 2$ gives $U(x) = \frac{1}{A \alpha (2-\alpha)} x^{2-\alpha} + (\alpha-1)\ln x + C$, whose leading term at large $x$ is a power-law. Setting $\tau(x)=A x^2$ we get $U(x) = \left(1+\frac{1}{2A} \right) \ln x + C$, consistent with the logarithmic example above upon identifying $U_0=1+1/(2A)$. Both satisfy the boundary condition at infinity.

Setting $\tau(x)=x^3$ and integrating gives $U(x)=2\ln x-\frac{1}{3x}+C$, but the boundary condition $\lim_{x \to \infty} e^{-U(x)}\tau'(x)=0$ is not obeyed, indicating a problem. Indeed, plugging this potential back into \cref{eq:integral_MFPT} does not give the prescribed $\tau(x)$, but rather $\tau'(x)=3x^2\left(1-e^{-1/3x}\right)$, such that $\tau(x)\simeq x^2/2$ for large $x$ (End Matter). This highlights that not all functions $\tau(x)$ are realizable as an MFPT of an overdamped Brownian particle in a potential. Next, we find the exact condition for $\tau(x)$ to be admissible.

\textit{The admissibility condition.---}A necessary condition follows by differentiating \cref{eq:integral_MFPT}. Writing $\mathcal Z(x)=\int_x^\infty dz\,e^{-U(z)}$, one has $\tau'(x)=e^{U(x)}\mathcal Z(x)$. Since $\mathcal Z'(x)=-e^{-U(x)}$, this gives $\tau'(x)=-\mathcal Z(x)/\mathcal Z'(x)$, or equivalently $d\ln\mathcal Z(x)/dx=-1/\tau'(x)$. Hence, for any finite $x$ and $R>x$, $\mathcal Z(R)=\mathcal Z(x)\exp\left[-\int_x^R dy/\tau'(y)\right]$. Now, combining $\mathcal Z(x)=e^{-U(x)}\tau'(x)$ with the boundary condition at infinity in \cref{eq:BCs}, we get $\lim_{R\to\infty}\mathcal{Z}(R)=0$. Since $\mathcal Z(x)$ is finite and nonzero for finite $x$, this is possible only if \begin{equation}\label{eq:integral_condition}
    \int^\infty \frac{dx}{\tau'(x)}=\infty .
\end{equation}
In particular, for the trial choice $\tau(x)=x^3$, one has $\tau'(x)=3x^2$, so $\int^\infty dx/\tau'(x)\sim\int^\infty dx/x^2$ converges. Hence, the cubic profile violates \cref{eq:integral_condition} and cannot be a physical MFPT.

In the End Matter, we show that the admissibility condition in \cref{eq:integral_condition}, together with the trivial conditions that $\tau(0)=0$, that $\tau(x)$ is twice continuously differentiable, and that $\tau'(x)>0$, are also sufficient. Thus, we provide necessary and sufficient conditions for the existence of a landscape that realizes $\tau(x)$ as the MFPT.  

\textit{Physical interpretation of the admissibility condition.---}Note, that $\mathcal{Z}(x)$ is the partition function of an ideal gas held on the ray $(x,\infty)$. The boundary condition in \cref{eq:BCs} is equivalent to $\mathcal{Z}(x)<\infty$, i.e., to the existence of an equilibrium state on that ray.

The inverse of $\tau'$ is equal to $-d\ln \mathcal{Z}(x)/dx$, which is also the force per particle applied by such a gas on a wall at $x$. Thus, \cref{eq:integral_condition} implies that the work required to compress the gas grows indefinitely as the wall is pushed to infinity. If this is not the case, one can conclude that no equilibrium state can be established on the ray, and therefore the MFPT diverges.

\textit{No superquadratic growth for eventually convex MFPT profiles.---}For a pure power law $\tau(x)=Ax^\alpha$, the realizability criterion immediately gives
\begin{equation}
    \int^\infty \frac{dx}{\tau'(x)}
    =
    \frac{1}{A\alpha}\int^\infty dx\,x^{1-\alpha},
\end{equation}
which converges for $\alpha>2$. Hence, no pure superquadratic power law is realizable. 

We now extend this conclusion to profiles with \textit{log-log asymptotics with exponent $\alpha$}, i.e., that for every $\epsilon>0$ there exist $X>0$ such that for any $x>X$
\begin{equation}\label{eq:loglog}
    A^{1-\epsilon}x^{\alpha-\epsilon}\leq\tau(x)\leq A^{1+\epsilon}x^{\alpha+\epsilon}.
\end{equation}
Because this asymptotic relation does not by itself determine the behavior of $\tau'(x)$, we first investigate the case where the slope is eventually nondecreasing, or equivalently where $\tau(x)$ is eventually convex. 

Thus, suppose that $A>0$, $\alpha>2$, and that there exists $X_0$ such that $\tau'(x)$ is nondecreasing for $x>X_0$. Then, for $x>X_0$,
\begin{equation}
    \tau(x)-\tau(X_0)
    =
    \int_{X_0}^x dy\,\tau'(y)
    \leq
    (x-X_0)\tau'(x),
\end{equation}
where the inequality follows from the monotonicity of $\tau'(x)$ on this interval. Since $\tau(X_0)$ is fixed while $\tau(x)$ obeys \cref{eq:loglog}, there exists $X'>X_0$ such that, for all $x>X'$, $\tau(x)-\tau(X_0)\geq A^{1-\epsilon}x^{\alpha-\epsilon}$ for $0<\epsilon<\alpha-2$. Combining the two bounds gives
\begin{equation}
    \tau'(x)
    \geq
   A^{1-\epsilon}\frac{x^{\alpha-\epsilon}}{x-X_0}
    \geq
    A^{1-\epsilon}x^{\alpha-1-\epsilon},
    \qquad x>X' .
\end{equation}
Therefore,
\begin{equation}
    \int_{X'}^\infty \frac{dx}{\tau'(x)}
    \leq
    A^{\epsilon-1}
    \int_{X'}^\infty dx\,x^{1-\alpha+\epsilon}
    <\infty,
\end{equation}
since $\alpha>2$ and $0<\epsilon<\alpha-2$, which contradicts the realizability condition \cref{eq:integral_condition}. Hence, no admissible potential can produce an eventually convex MFPT with log-log asymptotics of exponent $\alpha>2$.

The conclusion is that quadratic growth is the largest power-law scaling compatible with an eventually convex MFPT profile, i.e., with $\tau'(x)$ eventually increasing. If a realizable MFPT nevertheless exhibits superquadratic growth on logarithmic scales, then $\tau'(x)$ cannot be eventually monotone. Indeed, an eventually decreasing positive $\tau'(x)$ would give at most linear growth of $\tau(x)$, whereas an eventually increasing $\tau'(x)$ is ruled out by the argument above. Thus $\tau'$ must have local extrema arbitrarily far from the target. Since the force is continuous, \cref{eq:back_Kolmogorov} implies that $\tau''(x)$ is continuous, so these changes of monotonicity occur through genuine local minima and maxima of $\tau'(x)$.

The regime of motion corresponding to $\tau(x)$ growing with $x$ slower than quadratically practically corresponds to deterministic sliding of the particle towards the origin in potentials that grow sufficiently fast. To see this, one can approximate the inner integral in \cref{eq:integral_MFPT} using the Laplace method:
$\int_y^\infty dz \, e^{-U(z)} \simeq e^{-U(y)} \int_y^\infty dz \, e^{-U'(y)(z-y)} = \frac{e^{-U(y)}}{|F(y)|}$.
Substituting this into \cref{eq:integral_MFPT} one has
\begin{equation}
\tau(x) \simeq\int_0^x \frac{dy}{|F(y)|},
\label{eq:DD}
\end{equation}
the same as one would get from the solution of the deterministic overdamped equation of motion $\frac{dx}{dt} = -|F(x)|$ [\cref{eq:Langevin} without the random force term] for the time spent on the interval. For the power-law potentials $U(x) = \frac{1}{A \alpha (2-\alpha)} x^{2-\alpha}$, we get $\tau(x) = A x^\alpha$, which agrees with the result for overdamped Brownian motion to leading order. Thus, for $\alpha < 2$ the behavior is dominated by the deterministic motion. The Laplace's method approximation error is controlled by $|U''|/U'^2\sim x^{\alpha-2}$, which vanishes at large $x$ precisely when $\alpha<2$ and becomes of order unity at $\alpha=2$. Logarithmic potentials, $U(x)=\left(1+\frac{1}{2A}\right)\ln x$, mark the borderline case $\alpha=2$. For this potential, the deterministic sliding approximation \eqref{eq:DD} gives $\tau_{\mathrm{det}}(x)=A x^2/(1+2A)$, whereas the full stochastic dynamics gives the exact result $\tau(x)=Ax^2$. This discrepancy persists even at large $x$, indicating substantial contribution of thermal fluctuations to the MFPT.

\textit{Superquadratic scaling requires potential wells arbitrarily far from
the target.---}The preceding discussion suggests a simple physical distinction. In a monotone landscape, the particle moves toward the target without encountering local traps, whereas potential wells delay its progress through repeated trapping and escape. This raises a natural question: can superquadratic MFPT growth occur in a landscape that is eventually free of wells, or is trapping unavoidable? To answer this question, we now formulate the growth law without assuming monotonic behavior of $\tau'(x)$.

In the End Matter, we prove the following: if the landscape has no local minimum beyond some distance $X$, then $\tau(x)$ cannot have a log-log exponent $\alpha>2$. Admissibility forces $U'(x)\geq0$ far away, since a landscape that eventually decreases makes $\mathcal{Z}$ diverge and the MFPT infinite; by \cref{eq:PotTau} this means $\tau''(x)\geq-1$, so $g(x)=\tau'(x)+x$ is non-decreasing. Consequently $\tau(x)-\tau(X)=\int_X^x[g(y)-y]\,dy\leq x\,\tau'(x)+x^2$, which for $\alpha>2$ bounds $\tau'$ from below by $\sim x^{\alpha-\epsilon-1}$ and renders $\int^\infty dx/\tau'$ convergent, contradicting \cref{eq:integral_condition}.

This is the central structural result of this Letter, and it is strictly stronger than the no-go above: convex profiles are a special case, since $\tau''\geq0$ implies $\tau''\geq-1$. The guaranteed existence of local minima in the potential, if $\tau(x)$ exhibits a log-log exponent $\alpha>2$, is a qualitative difference from the case of $\alpha\leq 2$, where a monotonic potential can produce such asymptotics. These results can serve as a tool to learn about the properties of a potential by measuring $\tau(x)$. If one observes a log-log exponent $\alpha>2$, one can conclude that the potential giving rise to such behavior must have local minima, resulting in trapping-dominated transport. This enriches existing tools for inferring potential landscape features from first-passage and transition path time statistics \cite{Thorneywork2020, Satija2020}.

We now give an example of an MFPT with superquadratic log-log asymptotics, and find the potential realizing it. Consider the case where 
\begin{equation}\label{eq:example}
    \frac{d\tau}{dx}=e^{\left(\alpha-1\right)\ln x\sin2\pi x}
    =x^{(\alpha-1)\sin 2\pi x} .
\end{equation}
At large $x$, $\tau(x)$ with this derivative has log-log exponent $\alpha$. We can see that this derivative is always positive, so to check if this MFPT is realizable, all that is needed is to check whether the condition in \cref{eq:integral_condition} holds, namely, that this integral diverges, \begin{equation}\label{eq:integral_condition_example}
    \int_0^\infty\frac{dx}{\tau'(x)}=\int_0^\infty e^{-\left(\alpha-1\right)\ln x\sin2\pi x}\,dx.
\end{equation}
In the End Matter, we bound this integral from below by a simple integral, and show that the latter diverges. The corresponding landscape follows from \cref{eq:Uquadrature}, and is given by
\begin{equation}\label{eq:Uexample}
    U(x)=(\alpha-1)\ln x\,\sin 2\pi x+\int^x dy\,y^{-(\alpha-1)\sin2\pi y} .
\end{equation}

\begin{figure}[t!]
\centering
\includegraphics[width=0.45\textwidth]{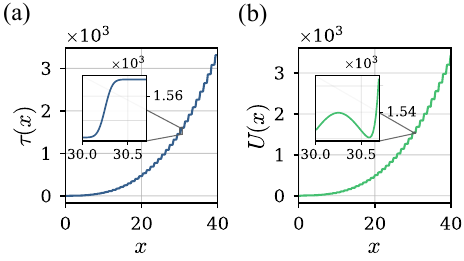}
  \caption{\textbf{Superquadratic MFPT scaling can be obtained via consecutive trapping.} (a) the MFPT obtained by integrating \cref{eq:example} with $\alpha=3$. (b) The corresponding energy landscape, given by \cref{eq:Uexample}.}
  \label{fig:Example}
\end{figure}

The resulting MFPT and potential for $\alpha=3$ are shown in Fig.~\ref{fig:Example}. The landscape consists of one well per unit cell. Each well is bounded on the left by a barrier of height $\Delta U_L(x)\sim(\alpha-1)\ln x$ and on the right by a cliff of height $\sim x^{\alpha-1}/\sqrt{\ln x}$, which makes backward hops negligible. The first passage thus reduces to a sequence of Arrhenius escape problems, one per well. Escaping the $n$-th well takes $t_n\sim e^{\Delta U_L(n)}=n^{\alpha-1}$, so logarithmically growing barriers yield power-law trapping times. Summing over the $\sim x$ wells gives $\tau(x)\sim\sum_{n\le x}n^{\alpha-1}\sim x^{\alpha}$. Each step of the staircase in Fig.~\ref{fig:Example}(a) is one such waiting time, as in a continuous-time random walk~\cite{Montroll1965, Metzler2000}. Details are in the Supplemental Material \cite{SM}; the End Matter confirms $\alpha=3$ on a log-log plot.

\textit{Conclusions---}To summarize, we have solved the inverse mean first-passage time problem for a one-dimensional overdamped particle: the landscape realizing a prescribed MFPT profile is unique and explicit, and admissible precisely when $\int^\infty dx/\tau'$ diverges. The criterion separates two transport mechanisms. In monotone landscapes, the particle slides, the MFPT cannot grow faster than quadratically with the starting distance, and the deterministic estimate \cref{eq:DD} applies. Beyond quadratic growth, transport must be trapping-controlled, and the landscape necessarily contains wells arbitrarily far from the target.

The practical consequence is a landscape diagnostic that requires no knowledge of the landscape itself. Measuring $\tau(x)$ and observing an effective exponent larger than $\alpha=2$ suffices to conclude that the underlying potential is non-monotone, without resolving it directly. 

Finally, since the potential landscape fixes the full dynamics, it also determines the entire first passage time distribution $\psi(t|x)$, for a given initial position $x$. Thus, an admissible $\tau(x)$, which determines the potential, also determines $\psi(t|x)$. Measuring $\tau(x)$ thus predicts the full first-passage statistics, and comparing this prediction with measured distributions tests whether the dynamics is indeed a one-dimensional overdamped diffusion in a static landscape.

\begin{acknowledgments}
This project has received funding from the European Research Council (ERC) under the European Union’s Horizon 2020 research and innovation program (Grant Agreement No. 947731). T.D.K. is supported by the Adams Fellowships Program of The Israel Academy of Sciences and Humanities.
\end{acknowledgments}

\bibliography{bibliography}

\newpage
\section*{End Matter}

\textit{Appendix A: The cubic profile and the landscape it generates---}For $\tau(x)=x^3$, \cref{eq:PotTau} gives $U'(x)=(1+6x)/(3x^2)$ and hence
$U(x)=2\ln x-1/(3x)$ up to a constant. The tail integral can be evaluated in
closed form,
\begin{equation}
    \mathcal Z(x)=\int_x^\infty \frac{e^{\frac{1}{3y}}}{y^{2}}\,dy=3\left(e^{\frac{1}{3x}}-1\right),
\end{equation}
so that the MFPT actually generated by this landscape has
\begin{equation}
\begin{split}
    \tau_U'(x)&=e^{U(x)}\mathcal Z(x)=3x^{2}\left(1-e^{-\frac{1}{3x}}\right)\\
    &\xrightarrow[x\to\infty]{} x-\frac16+\mathcal O(x^{-1}),
\end{split}
\end{equation}
i.e. $\tau_U(x)\simeq x^2/2$ rather than the prescribed $x^3$. Consistently, the
boundary condition at infinity fails for the prescribed profile:
$e^{-U(x)}\tau'(x)=3e^{1/3x}\to3\neq0$.\\

\textit{Appendix B: The realizability criterion is necessary and sufficient---}Let $\tau(x)$ be twice continuously differentiable on $[0,\infty)$, with
$\tau(0)=0$, $\tau'(x)>0$, and $\tau$ finite on compacts. We claim that $\tau(x)$ is
the MFPT of some admissible landscape if and only if
$\int^\infty dx/\tau'(x)=\infty$, and that the landscape is then uniquely given by \cref{eq:Uquadrature} up to an
additive constant.

Necessity was shown in the main text. For sufficiency, fix $x_0>0$, define
\begin{equation}
    W(x)=\int_{x_0}^{x}\frac{dy}{\tau'(y)},
    \qquad
    U(x)=\ln\tau'(x)+W(x),
\end{equation}
and note that $U$ is continuously differentiable with
$U'=(1+\tau'')/\tau'$ being continuous, so the force is continuous as required. Since
$W'(x)=1/\tau'(x)$,
\begin{equation}
    e^{-U(x)}=\frac{e^{-W(x)}}{\tau'(x)}=W'(x)e^{-W(x)}=-\frac{d}{dx}e^{-W(x)} ,
\end{equation}
and therefore, for any $R>x$,
\begin{equation}
    \int_x^{R} dy\,e^{-U(y)}=e^{-W(x)}-e^{-W(R)} .
\end{equation}
If $\int^\infty dy/\tau'=\infty$, then $W(R)\to\infty$ and
$\mathcal{Z}(x)=e^{-W(x)}$, which is finite and vanishes at infinity, so the
boundary condition at infinity holds. Consequently
\begin{equation}
    e^{U(x)}\mathcal Z(x)=\tau'(x)e^{W(x)}e^{-W(x)}=\tau'(x) ,
\end{equation}
and integrating from $0$ to $x$ with $\tau(0)=0$ returns the prescribed profile,
which proves sufficiency.\\

\textit{Appendix C: Monotone landscapes cannot support superquadratic growth---}Suppose $\tau(x)$ is realizable and that the potential landscape has no local minimum on
$(X,\infty)$. We first argue that $U'(x)\geq0$ there. Since $U'$ is continuous and
has no sign change from negative to positive on $(X,\infty)$, either $U'(x)\geq0$
throughout, or there is a point beyond which $U'(x)\leq0$. The latter is excluded by
admissibility: if $U(x)$ is eventually non-increasing then $e^{-U(x)}$ is eventually
non-decreasing, $\mathcal Z(x)=\int_x^\infty e^{-U(x)}$ diverges, and the MFPT is
infinite. Hence $U'(x)\geq0$ on $(X,\infty)$ and, by \cref{eq:PotTau} together with
$\tau'(x)>0$,
\begin{equation}
    \tau''(x)\geq-1,
    \qquad x>X .
\end{equation}

Define $g(x)=\tau'(x)+x\geq 0$. Then $g'(x)=\tau''(x)+1\geq0$, so $g(x)$ is non-decreasing on
$(X,\infty)$, and for $x>X$
\begin{equation}
\begin{split}
    \tau(x)-\tau(X)&=\int_X^x \left[g(y)-y\right]dy\\
    &\leq (x-X)g(x)-\frac{x^2-X^2}{2}
    \leq x\,\tau'(x)+x^{2} .
\end{split}
\end{equation}
Using it, we can create a lower bound on $\tau'(x)$
\begin{equation}
    \tau'(x)\geq \frac{\tau(x)-\tau(X)}{x}-x.
\end{equation}

Assume by contradiction that $\tau(x)$ has log-log exponent $\alpha>2$ in the
sense of \cref{eq:loglog}, and fix $0<\epsilon<\alpha-2$. By \cref{eq:loglog},
there exists $X_1>X$ such that $\tau(x)\geq A^{1-\epsilon}x^{\alpha-\epsilon}$ for all $x>X_1$. Since $\tau(X)$ is a fixed constant, the lower bound above gives
\begin{equation}
    \tau'(x)\geq A^{1-\epsilon}x^{\alpha-1-\epsilon}-\frac{\tau(X)}{x}-x,
    \qquad x>X_1 .
\end{equation}
Because $\alpha-1-\epsilon>1$, the first term dominates the other two at large $x$, so there exists $X_2\geq X_1$ such that
\begin{equation}
    \tau'(x)\geq \frac{A^{1-\epsilon}}{2}\,x^{\alpha-1-\epsilon},
    \qquad x>X_2 .
\end{equation}
Therefore,
\begin{equation}
    \int_{X_2}^\infty\frac{dx}{\tau'(x)}\leq 2A^{\epsilon-1}\int_{X_2}^\infty dx\,x^{1-\alpha+\epsilon}<\infty,
\end{equation}
since $1-\alpha+\epsilon<-1$. This contradicts the realizability condition \cref{eq:integral_condition} and the initial assumption that $\tau(x)$ is realizable.

\textit{Appendix D: Divergence of the integral in \cref{eq:integral_condition_example}---}In this appendix, we show that the integral in
\cref{eq:integral_condition_example} diverges. We do so by bounding it from below by a diverging integral. We denote the integral in \cref{eq:integral_condition_example} by $J$ and take
$\alpha>2$. Since the integrand is positive,
\begin{equation}
    J\geq\int_1^\infty e^{-(\alpha-1)\ln x\sin2\pi x}\,dx .
\end{equation}
We integrate over intervals $[n,n+1]$ and sum over the integers $n$, further
shrinking each integration domain to $[n+7/12,\,n+11/12]$, which yields
\begin{equation}
    J\geq\sum_{n=1}^{\infty}\int_{n+\frac{7}{12}}^{n+\frac{11}{12}} e^{-(\alpha-1)\ln x\sin2\pi x}\,dx .
\end{equation}
On each of these intervals $-\sin 2\pi x\geq1/2$ and $\ln x\geq\ln n$,
so that $-(\alpha-1)\ln x\sin2\pi x\geq\frac{\alpha-1}{2}\ln n$. Hence
\begin{equation}
    J\geq\sum_{n=1}^{\infty}\int_{n+\frac{7}{12}}^{n+\frac{11}{12}} e^{\frac{\alpha-1}{2}\ln n}\,dx
    =\frac13\sum_{n=1}^\infty n^{\frac{\alpha-1}{2}} ,
\end{equation}
where $1/3$ is the width of the shrunken interval. This sum diverges for the superquadratic case $\alpha>2$ of interest. Thus the integral $J$ diverges, and the profile of \cref{eq:example} is realizable.

\textit{Appendix E: log-log plot of the MFPT constructed from \cref{eq:example}---}
\begin{figure}[h]
    \centering
    \includegraphics[width=0.85\linewidth]{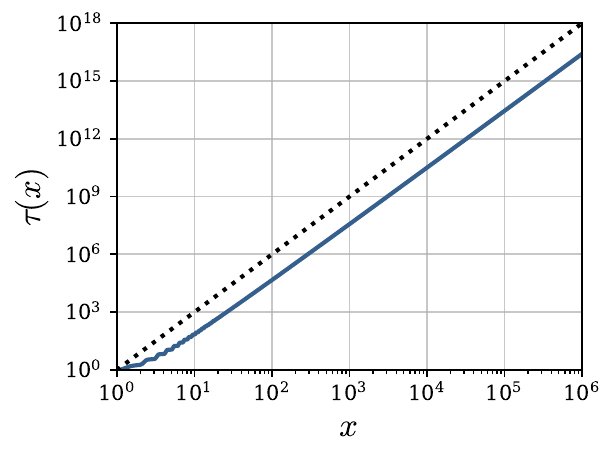}
    \caption{The solid blue line is the MFPT for the example presented in the text, obtained by numerical integration of \cref{eq:example} with $\alpha=3$. The dotted black line is $x^3$. The fact that the two are parallel shows that indeed $\tau(x)$ exhibits a log-log asymptotics with $\alpha=3$.}
    \label{fig:log-log}
\end{figure}

\subfile{SM}
\end{document}

%% file: SM.tex
\clearpage
\appendix
\setcounter{equation}{0}
\setcounter{figure}{0}
\setcounter{table}{0}
\setcounter{section}{0}
\renewcommand{\theequation}{S\arabic{equation}}
\renewcommand{\thefigure}{S\arabic{figure}}
\renewcommand{\thetable}{S\arabic{table}}
\renewcommand{\thesection}{S\arabic{section}}

\let\addcontentsline\origaddcontentsline

\ifSubfilesClassLoaded{%
  \title{{\Large Supplemental Material}\\[6pt]
  The Inverse Mean First-Passage Time Problem}
  \author{Tommer D. Keidar}
  \affiliation{School of Chemistry, The Center for Computational Molecular and Materials Science, The Center for Physics and Chemistry of Living Systems, Tel Aviv University, Tel Aviv, Israel}
  \author{Shlomi Reuveni}
  \email{Contact author: shlomire@tauex.tau.ac.il}
  \affiliation{School of Chemistry, The Center for Computational Molecular and Materials Science, The Center for Physics and Chemistry of Living Systems, Tel Aviv University, Tel Aviv, Israel}
  \author{Igor M. Sokolov}
  \email{Contact author: sokolov@physik.hu-berlin.de}
  \affiliation{Institut f\"ur Physik, Humboldt-Universit\"at zu Berlin, Newtonstra{\ss}e 15, D-12489 Berlin, Germany}
  \date{\today}
  \maketitle
  \onecolumngrid
}{%
  \onecolumngrid
  \begin{center}
    {\Large\bfseries Supplemental Material}\\[6pt]
    {\large\bfseries The Inverse Mean First-Passage Time Problem}\\[10pt]
    Tommer D. Keidar$^1$, Shlomi Reuveni$^1$, and Igor M. Sokolov$^2$\\[4pt]
    {\small $^1$School of Chemistry, The Center for Computational Molecular and Materials Science, The Center for Physics and Chemistry of Living Systems, Tel Aviv University, Tel Aviv, Israel\\
    $^2$Institut f\"ur Physik, Humboldt-Universit\"at zu Berlin, Newtonstra{\ss}e 15, D-12489 Berlin, Germany}
  \end{center}
}

\tableofcontents

\section{Deriving Eq.~(3) and the boundary condition in Eq.~(4) using the flux-over-population approach}

To derive Eq.~(3) of the main text and the appropriate boundary
conditions for the backward Kolmogorov equation, we will consider the case where particles are injected at rate $I$ at
some point $x_0>0$, and where there is an absorbing boundary at $x=0$. At the
steady state, the mean number of particles in the system $\langle n\rangle$ is
equal to $I \tau(x_0)$. This stems from Little's
law~\cite{Little1961}. Intuitively, if particles are injected at a higher
rate or leave the system more slowly, we expect their number to increase. Using the
continuity equation for the flux of the particles $j$, we get that at the steady
state
\begin{equation}
    -\frac{dj(x)}{dx}+I\delta(x-x_0)=0,
\end{equation}
and the flux obeys
\begin{equation}
    j(x)=-\frac{dU}{dx}n(x)-\frac{dn}{dx}.
\end{equation}
In the steady state, we should not have flux from $x_0$ to infinity; otherwise
some particles will escape to infinity instead of leaving from $x=0$. Therefore,
the flux $j(x)$ is equal to $-I$ for $x<x_0$ and $0$ for $x>x_0$. So we get that
\begin{equation}
    \begin{cases}
        \frac{dn}{dx}+n(x)\frac{dU}{dx}=I&\text{for $x<x_0$,}\\[2pt]
        \frac{dn}{dx}+n(x)\frac{dU}{dx}=0&\text{for $x>x_0$.}
    \end{cases}
\end{equation}
Solving these equations, we get
\begin{equation}
   n(x)= \begin{cases}
        e^{-U(x)}\left(C_1+I\int_0^xe^{U(y)}\,dy\right)&\text{for $x<x_0$,}\\[2pt]
        C_2e^{-U(x)}&\text{for $x>x_0$.}
    \end{cases}
\end{equation}
At $x=0$ there is an absorbing boundary so $n(0)=0$. At $x=x_0$ the density of particles should be continuous. This is because discontinuities result in diverging fluxes and thus are not stable. So we get that $C_1=0$ and $C_2=I\int_0^{x_0}e^{U(y)}\,dy$.

The MFPT from $x_0$ is thus
\begin{equation}
    \tau(x_0)=\frac{\langle n\rangle}{I}=\frac{\int_0^\infty n(x)\,dx}{I}=\int_0^{x_0}dx\,e^{-U(x)}\int_0^xdy\,e^{U(y)}+\int_0^{x_0}dy\,e^{U(y)}\int_{x_0}^\infty dx\,e^{-U(x)}.
\end{equation}
Changing the order of integration in the first term, we arrive at Eq. (3) in the main text.

From this derivation, it is clear why we need to require that
$\mathcal{Z}(x_0)<\infty$. Otherwise, the mean number of particles in the system will diverge and therefore the MFPT too. 

\section{Solving the backward equation for the MFPT}

We consider the mean first-passage time $\tau(x)$ to reach an absorbing boundary
at $x=0$ for a one-dimensional overdamped diffusion on the half-line
$x\in[0,\infty)$. In units where $D=\mu k_B T=1$, the backward equation is
\begin{equation}
    \tau''(x)-U'(x)\tau'(x)=-1,
    \qquad x>0 ,
\end{equation}
and the absorbing boundary condition is $\tau(0)=0$.

The equation is second order in $\tau(x)$, but first order in $\tau'(x)$. We
therefore define $p(x)=\tau'(x)$, such that
\begin{equation}
    p'(x)-U'(x)p(x)=-1 .
\end{equation}
This is a first-order linear equation for $p(x)$. Multiplying by the integrating
factor $e^{-U(x)}$, the left-hand side becomes a total derivative, and
\begin{equation}\label{eq:IntegratingFactor}
    \frac{d}{dx}\left[e^{-U(x)}p(x)\right]
    =
    -e^{-U(x)} .
\end{equation}

To solve this equation on the ray $x\in[0,\infty)$, we will start by solving it on the finite interval $x\in[0,L]$, with a reflecting boundary at $x=L$, and denote this solution by $\tau_L(x)$ and $\tau'_L(x)=p_L(x)$. Then, we will take the limit $L\to\infty$. The reflecting boundary sets $p_L(L)=0$. Using \cref{eq:IntegratingFactor} and this boundary condition, we get
\begin{equation}
    p_L(x)=\tau_L'(x)
    =
    e^{U(x)}
    \int_x^L dy\,e^{-U(y)} .
\end{equation}
Integrating once more from $0$ to $x$, and using the absorbing boundary condition at $x=0$, $\tau_L(0)=0$, gives
\begin{equation}
    \tau_L(x)
    =
    \int_0^x dz\, e^{U(z)}
    \int_z^L dy\,e^{-U(y)} .
\end{equation}

The half-line result is obtained by taking the limit $L\to\infty$. If
$\mathcal Z(x)=\int_x^\infty dy\,e^{-U(y)}<\infty$, then
\begin{equation}
    \tau'(x)
    =
    e^{U(x)}
    \int_x^\infty dy\,e^{-U(y)} ,
\end{equation}
and hence
\begin{equation}
    \tau(x)
    =
    \int_0^x dz\, e^{U(z)}
    \int_z^\infty dy\,e^{-U(y)} .
\end{equation}

The corresponding boundary condition at infinity is not a condition on $\tau(x)$
itself, since $\tau(x)$ may diverge as $x\to\infty$. Rather, the natural boundary
condition is $\lim_{x\to\infty} e^{-U(x)}\tau'(x)=0$. Indeed, from the expression
above, $e^{-U(x)}\tau'(x)=\int_x^\infty dy\,e^{-U(y)}$, and the right-hand side
vanishes as $x\to\infty$ whenever the tail integral is finite. Thus the backward
problem on the half-line should be written as
\begin{equation}
\begin{aligned}
    \tau''(x)-U'(x)\tau'(x)&=-1,
    \qquad x>0,\\
    \tau(0)&=0,\\
    \lim_{x\to\infty}e^{-U(x)}\tau'(x)&=0 .
\end{aligned}
\end{equation}

Equivalently, without imposing the boundary condition at infinity, the first
integration gives $e^{-U(x)}\tau'(x)=C-\int^x dy\,e^{-U(y)}$, where $C$ is an
integration constant. The physical MFPT is selected by choosing $C$ such that
$e^{-U(x)}\tau'(x)\to0$ as $x\to\infty$, which recovers the formula above.

\subsection{Derivation using two absorbing boundaries}

We can also consider the case of a particle diffusing in the box $[0,L]$, where both boundaries are absorbing. This translates to the boundary conditions
$\tau(0)=\tau(L)=0$. Then, by taking the limit $L\to\infty$, we should get back
Eq. (3) in the main text. Using the same substitution and integrating factor as
above, we get
\begin{equation}
    \tau'(x)=e^{U(x)}\left(A-\int_0^xe^{-U(y)}\,dy\right),
\end{equation}
where $A$ is some constant. Integrating again, we get that
\begin{equation}
    \tau(x)=B+\int_0^x dy\,e^{U(y)}\left(A-\int_0^y dz\,e^{-U(z)}\right).
\end{equation}
Using the boundary condition at $x=0$, we get that $B=0$. Using the boundary
condition at $x=L$, we get
\begin{equation}
    A=\frac{\int_0^Lds\,e^{U(s)}\int_0^sdz\,e^{-U(z)}}{\int_0^Lds\,e^{U(s)}},
\end{equation}
resulting in
\begin{equation}
    \tau(x)=\int_0^xdy\,e^{U(y)}\frac{\int_0^L ds\,e^{ U(s)}\int_y^sdz\,e^{-U(z)}}{\int_0^Lds\,e^{U(s)}}.
\end{equation}
The limit as $L\to\infty$ can be taken using L'H\^opital's rule, which gives
\begin{equation}
    \tau(x)=\int_0^xdy\,e^{U(y)}\lim_{L\to\infty}\int_y^Ldz\,e^{-U(z)}.
\end{equation}
The above equation diverges when the inner integral diverges, and converges to Eq. (3) in the main text if this integral converges.

\section{Continuous-time random walk analysis of the example illustrated in Fig. 2} 

Recalling the potential given in the main text
\begin{equation}
    U(x)=(\alpha-1)\ln x\,\sin 2\pi x+\int^x dy\,e^{-(\alpha-1)\ln y\sin2\pi y}.
\end{equation}
We will write $x$ as $n+z$, where $n\equiv\lfloor x\rfloor$, and $z$ is its fractional part. So the potential can be written as 
\begin{equation}
\begin{split}
    U(n,z)=&U(n)+(\alpha-1)\ln{(n+z)}\sin 2\pi z+\int_0^zdy\,e^{-(\alpha-1)\ln{(n+y)}\sin2\pi y}.
\end{split}
\end{equation}
Taking the first two derivatives with respect to $z$, we get
\begin{equation}
    \begin{split}
        \frac{\partial U(n,z)}{\partial z}=&\frac{\alpha-1}{n+z}\sin2\pi z+2\pi(\alpha-1)\ln(n+z)\cos2\pi z+e^{-(\alpha-1)\ln(n+z)\sin2\pi z}\simeq \\
        &\frac{\alpha-1}{n}\sin2\pi z+2\pi(\alpha-1)\ln(n)\cos2\pi z+e^{-(\alpha-1)\ln(n)\sin2\pi z},\\
        \frac{\partial^2U(n,z)}{\partial z^2}=&4\pi\frac{\alpha-1}{n+z}\cos2\pi z-(\alpha-1)\left(\frac{1}{(n+z)^2}+4\pi^2\ln(n+z)\right)\sin2\pi z-\\
        &\left(\frac{\alpha-1}{n+z}\sin2\pi z+2\pi(\alpha-1)\ln(n+z)\cos2\pi z\right)e^{-(\alpha-1)\ln(n+z)\sin2\pi z}\simeq\\
        &4\pi\frac{\alpha-1}{n}\cos2\pi z-4\pi^2(\alpha-1)\ln(n)\sin2\pi z-\left(\frac{\alpha-1}{n}\sin2\pi z+2\pi(\alpha-1)\ln(n)\cos2\pi z\right)e^{-(\alpha-1)\ln(n)\sin2\pi z},
    \end{split}
\end{equation}
where we used the fact that $x\gg1$, so $n\gg z$. The derivative has two zeros, one in the vicinity of $z=1/4$, and the other near $z=1/2$. Evaluating the derivative at $z=1/4+\epsilon$ we get
\begin{equation}
    0=\left.\frac{\partial U(n,z)}{\partial z}\right\rvert_{z=\frac{1}{4}+\epsilon}\simeq\frac{(\alpha-1)\cos2\pi\epsilon}{n}-2\pi(\alpha-1)\ln n\sin(2\pi \epsilon)+e^{-(\alpha-1)\ln n\cos2\pi\epsilon}\simeq\frac{\alpha-1}{n}-4\pi^2\epsilon(\alpha-1)\ln n+n^{1-\alpha}.
\end{equation}
For $\alpha>2$ we thus obtain
\begin{equation}
    \epsilon=\frac{1}{4\pi^2n\ln n}.
\end{equation}
The second derivative is negative there, so this is a maximum. Now, we evaluate the derivative at $z=1/2+\delta$ and equate it to zero
\begin{equation}
    0=\left.\frac{\partial U(n,z)}{\partial z}\right\rvert_{z=\frac{1}{2}+\delta}\simeq-2\pi(\alpha-1)\ln n\cos(2\pi \delta)+e^{(\alpha-1)\ln n\sin2\pi\delta}\simeq-2\pi(\alpha-1)\ln n+e^{2\pi(\alpha-1)\ln n\delta}.
\end{equation}
We therefore find
\begin{equation}\label{eq:delta_value}
    \delta=\frac{\ln\left[2\pi(\alpha-1)\ln n\right]}{2\pi(\alpha-1)\ln n}.
\end{equation}
The second derivative is positive at this point, so this is a minimum.

The potential energy at $z=1/4+\epsilon$ is
\begin{equation}
    U\left(n,\frac{1}{4}+\epsilon\right)\simeq U(n)+(\alpha-1)\ln n\cos(2\pi\epsilon)+\int_0^{\frac{1}{4}+\epsilon}dy\,e^{-(\alpha-1)\ln n\sin2\pi y}.
\end{equation}
The integrand is sharply peaked around the minimum of $\sin 2\pi y$, which is at $y=0$ in the domain of integration. We thus approximate the integral as
\begin{equation}\label{eq:Laplace_for_barrier}
    \int_0^{\frac{1}{4}+\epsilon}dy\,e^{-(\alpha-1)\ln n\sin2\pi y}\simeq \int_0^\infty dy\,e^{-2\pi (\alpha-1)y\ln  n}=\frac{1}{2\pi(\alpha-1)\ln n}.
\end{equation}
Overall,
\begin{equation}
    U\left(n,\frac{1}{4}+\epsilon\right)\simeq U(n)+(\alpha-1)\ln n.
\end{equation}

The potential energy at $z=1/2+\delta$ is
\begin{equation}
    U\left(n,\frac{1}{2}+\delta\right)\simeq U(n)-(\alpha-1)\ln n\sin 2\pi \delta+\int_0^{\frac{1}{2}+\delta}dy\,e^{-(\alpha-1)\ln n\sin2\pi y}.
\end{equation}
We can split the integration domain into three domains $[0,1/4),\,[1/4,1/2),\,[1/2,1/2+\delta)$. The first two evaluate to the same value as the integral in Eq. (\ref{eq:Laplace_for_barrier}). For the third domain, we get
\begin{equation}
\begin{split}
    \int_\frac{1}{2}^{\frac{1}{2}+\delta}dy\,e^{-(\alpha-1)\ln n\sin2\pi y}=&\int_0^\delta dy\,e^{(\alpha-1)\ln n\sin2\pi y}\simeq \\&\int_0^\delta dy\,e^{2\pi(\alpha-1)y\ln n}=\frac{e^{2\pi(\alpha-1)\delta\ln n}-1}{2\pi(\alpha-1)\ln n}=1-\frac{1}{2\pi(\alpha-1)\ln n},
\end{split}
\end{equation}
where we have used the value of $\delta$ given in Eq. (\ref{eq:delta_value}). Overall,
\begin{equation}
    U\left(n,\frac{1}{2}+\delta\right)\simeq U(n)-\ln\left[2\pi(\alpha-1)\ln n\right]+1.
\end{equation}

The energy barrier that needs to be crossed for the particle to move to the left, from the $n$-th trap to the $(n-1)$-th trap, is given by
\begin{equation}
    \Delta U_L(n)=U\left(n,\frac{1}{4}+\epsilon\right)- U\left(n,\frac{1}{2}+\delta\right)\simeq (\alpha-1)\ln n.
\end{equation}
For the move in the other direction, we get
\begin{equation}
    \Delta U_R(n)=U\left(n+1,\frac{1}{4}+\epsilon\right)- U\left(n,\frac{1}{2}+\delta\right)\simeq U(n+1)-U(n)+(\alpha-1)\ln n.
\end{equation}
The difference between $U(n+1)$ and $U(n)$ is given by
\begin{equation}
    U(n+1)-U(n)=\int_0^1dy\,e^{-(\alpha-1)\ln{(n+y)}\sin 2\pi y}\simeq\int_0^1dy\,e^{-(\alpha-1)\ln{n}\sin 2\pi y}=I_0((\alpha-1)\ln n),
\end{equation}
utilizing the integral representation of the modified Bessel functions of the first kind $I_\nu(u)$. Recalling its asymptotic behavior,
\begin{equation}
    I_0(u)\simeq\frac{e^u}{\sqrt{2\pi u}},
\end{equation}
we get
\begin{equation}
    U(n+1)-U(n)\simeq \frac{n^{\alpha-1}}{\sqrt{2\pi(\alpha-1)\ln n}}.
\end{equation}
We thus find that the height of the energy barrier for crossing from the $n$-th trap to the $(n+1)$-th is given by
\begin{equation}
    \Delta U_R(n)\simeq \frac{n^{\alpha-1}}{\sqrt{2\pi(\alpha-1)\ln n}}.
\end{equation}
The probability of crossing the barrier to the right before crossing the barrier to the left is given by
\begin{equation}
    p_R(n)=\frac{1}{1+e^{\Delta U_R(n)-\Delta U_L(n)}}\simeq e^{-\frac{n^{\alpha-1}}{\sqrt{2\pi(\alpha-1)\ln n}}}.
\end{equation}
As can be seen, this probability decays to zero quickly, so for large $x$ we can ignore it. 

The mean residence time in the $n$-th trap thus scales as 
\begin{equation}
    t_\text{trap}(n)\sim e^{\Delta U_L(n)}\sim n^{\alpha-1}.
\end{equation}

\noindent Concluding, we have found that the mean residence time scales as $n^{\alpha-1}$, and as the motion is almost unidirectional between traps, the MFPT is approximately the sum of the residence times in all traps between the initial point and the origin. This sum scales as $x^\alpha$, as required.

\ifSubfilesClassLoaded{\bibliography{bibliography}}{}

\end{document}

%% file: bibliography.bib
@article{scher2021unified,
  title={Unified approach to gated reactions on networks},
  author={Scher, Yuval and Reuveni, Shlomi},
  journal={Physical review letters},
  volume={127},
  number={1},
  pages={018301},
  year={2021},
  publisher={APS}
}

@article{Hanggi1990,
  title = {Reaction-rate theory: fifty years after {Kramers}},
  author = {H\"anggi, Peter and Talkner, Peter and Borkovec, Michal},
  journal = {Rev. Mod. Phys.},
  volume = {62},
  issue = {2},
  pages = {251--341},
  numpages = {0},
  year = {1990},
  month = {Apr},
  publisher = {American Physical Society},
  doi = {10.1103/RevModPhys.62.251},
  url = {https://link.aps.org/doi/10.1103/RevModPhys.62.251}
}

@article{pontryagin1933,
  author = {Pontryagin, L. and Andronov, A. and Witt, A.},
  title = {{Statistische Auffassung dynamischer Systeme}},
  journal = {Zh. Eksp. Teor. Fiz.},
  volume = {3},
  pages = {165--180},
  year = {1933}
}

@article{Kramers1940,
title = {Brownian motion in a field of force and the diffusion model of chemical reactions},
journal = {Physica},
volume = {7},
number = {4},
pages = {284-304},
year = {1940},
issn = {0031-8914},
doi = {https://doi.org/10.1016/S0031-8914(40)90098-2},
author = {H.A. Kramers}
}

@article{Little1961,
  title={A proof for the queuing formula: L= $\lambda$ W},
  author={Little, John DC},
  journal={Operations research},
  volume={9},
  number={3},
  pages={383--387},
  year={1961},
  publisher={INFORMS}
}

@article{Smoluchowski1918,
url = {https://doi.org/10.1515/zpch-1918-9209},
title = {{Versuch einer mathematischen Theorie der Koagulationskinetik kolloider Lösungen}},
author = {M. v. Smoluchowski},
pages = {129--168},
volume = {92U},
number = {1},
journal = {Zeitschrift für Physikalische Chemie},
doi = {doi:10.1515/zpch-1918-9209},
year = {1918}
}

@article{Farkas1927,
url = {https://doi.org/10.1515/zpch-1927-12513},
title = {{Keimbildungsgeschwindigkeit in übersättigten Dämpfen}},
author = {L. Farkas},
pages = {236--242},
volume = {125U},
number = {1},
journal = {Zeitschrift für Physikalische Chemie},
doi = {doi:10.1515/zpch-1927-12513},
year = {1927}
}

@book{Redner_2001, place={Cambridge}, title={A Guide to First-Passage Processes}, publisher={Cambridge University Press}, author={Redner, Sidney}, year={2001}}

@book{Metzler2014,
author = {Metzler, Ralf and Oshanin, Gleb and Redner, Sidney},
title = {First-Passage Phenomena and Their Applications},
publisher = {World Scientific},
year = {2014},
doi = {10.1142/9104},
address = {Singapore},
URL = {https://www.worldscientific.com/doi/abs/10.1142/9104}
}

@incollection{grebenkov2024,
  title={Target search problems},
  author={Grebenkov, Denis and Metzler, Ralf and Oshanin, Gleb},
  booktitle={Target Search Problems},
  pages={1--29},
  year={2024},
  publisher={Springer}
}

@article{Bray2013,
author = {Alan J. Bray and Satya N. Majumdar and Grégory Schehr},
title = {Persistence and first-passage properties in nonequilibrium systems},
journal = {Advances in Physics},
volume = {62},
number = {3},
pages = {225--361},
year = {2013},
publisher = {Taylor \& Francis},
doi = {10.1080/00018732.2013.803819},
URL = {https://doi.org/10.1080/00018732.2013.803819}
}

@article{Benichou2014,
title = {From first-passage times of random walks in confinement to geometry-controlled kinetics},
journal = {Physics Reports},
volume = {539},
number = {4},
pages = {225-284},
year = {2014},
issn = {0370-1573},
doi = {https://doi.org/10.1016/j.physrep.2014.02.003},
url = {https://www.sciencedirect.com/science/article/pii/S0370157314000118},
author = {O. Bénichou and R. Voituriez}
}

@article{berg1981,
  author = {Berg, O G and Winter, R B and von Hippel, P H},
  title = {Diffusion-driven mechanisms of protein translocation on nucleic acids. 1. Models and theory},
  journal = {Biochemistry},
  year = {1981},
  volume = {20},
  number = {24},
  pages = {6929--6948},
  doi = {10.1021/bi00527a028},
  issn = {0006-2960},
}

@article{gerstein1964,
  title={Random walk models for the spike activity of a single neuron},
  author={Gerstein, George L and Mandelbrot, Benoit},
  journal={Biophysical journal},
  volume={4},
  number={1},
  pages={41--68},
  year={1964},
  publisher={Elsevier}
}

@article{Ovaskainen2010,
title = {Stochastic models of population extinction},
journal = {Trends in Ecology \& Evolution},
volume = {25},
number = {11},
pages = {643-652},
year = {2010},
issn = {0169-5347},
doi = {https://doi.org/10.1016/j.tree.2010.07.009},
url = {https://www.sciencedirect.com/science/article/pii/S0169534710001801},
author = {Otso Ovaskainen and Baruch Meerson}
}

@inbook{Chicheportiche2014,
author = {Rémy Chicheportiche and Jean-Philippe Bouchaud},
title = {Some Applications of First-Passage Ideas to Finance},
booktitle = {First-Passage Phenomena and Their Applications},
chapter = {18},
year = {2014},
pages = {447-476},
doi = {10.1142/9789814590297_0018},
URL = {https://www.worldscientific.com/doi/abs/10.1142/9789814590297_0018}
}

@article{Majumdar2008,
author = {Satya N. Majumdar and Jean-Philippe Bouchaud},
title = {{Optimal time to sell a stock in the Black–Scholes model: comment on ‘Thou shalt buy and hold’, by A. Shiryaev, Z. Xu and X.Y. Zhou}},
journal = {Quantitative Finance},
volume = {8},
number = {8},
pages = {753--760},
year = {2008},
publisher = {Routledge},
doi = {10.1080/14697680802569093},
URL = {https://doi.org/10.1080/14697680802569093}
}

@article{Bray2000,
  title = {Random walks in logarithmic and power-law potentials, nonuniversal persistence, and vortex dynamics in the two-dimensional $\mathrm{XY}$ model},
  author = {Bray, A. J.},
  journal = {Phys. Rev. E},
  volume = {62},
  issue = {1},
  pages = {103--112},
  numpages = {0},
  year = {2000},
  month = {Jul},
  publisher = {American Physical Society},
  doi = {10.1103/PhysRevE.62.103},
  url = {https://link.aps.org/doi/10.1103/PhysRevE.62.103}
}

@article{Chupeau2020,
author = {Marie Chupeau  and Jannes Gladrow  and Alexei Chepelianskii  and Ulrich F. Keyser  and Emmanuel Trizac },
title = {Optimizing {Brownian} escape rates by potential shaping},
journal = {Proceedings of the National Academy of Sciences},
volume = {117},
number = {3},
pages = {1383-1388},
year = {2020},
doi = {10.1073/pnas.1910677116},
URL = {https://www.pnas.org/doi/abs/10.1073/pnas.1910677116}}

@article{Thorneywork2020,
author = {Alice L. Thorneywork  and Jannes Gladrow  and Yujia Qing  and Marc Rico-Pasto  and Felix Ritort  and Hagan Bayley  and Anatoly B. Kolomeisky  and Ulrich F. Keyser },
title = {Direct detection of molecular intermediates from first-passage times},
journal = {Science Advances},
volume = {6},
number = {18},
pages = {eaaz4642},
year = {2020},
doi = {10.1126/sciadv.aaz4642},
URL = {https://www.science.org/doi/abs/10.1126/sciadv.aaz4642}}

@article{Palyulin_2012,
doi = {10.1088/1742-5468/2012/03/L03001},
url = {https://doi.org/10.1088/1742-5468/2012/03/L03001},
year = {2012},
month = {mar},
publisher = {},
volume = {2012},
number = {03},
pages = {L03001},
author = {Palyulin, Vladimir V and Metzler, Ralf},
title = {How a finite potential barrier decreases the mean first-passage time},
journal = {Journal of Statistical Mechanics: Theory and Experiment}
}

@article{Zucca2009,
 ISSN = {10505164},
 URL = {http://www.jstor.org/stable/30243624},
 author = {Cristina Zucca and Laura Sacerdote},
 journal = {The Annals of Applied Probability},
 number = {4},
 pages = {1319--1346},
 publisher = {Institute of Mathematical Statistics},
 title = {On the Inverse First-Passage-Time Problem for a {Wiener} Process},
 volume = {19},
 year = {2009}
}

@article{chen2011,
  title={Existence and uniqueness of solutions to the inverse boundary crossing problem for diffusions},
  author={Chen, Xinfu and Cheng, Lan and Chadam, John and Saunders, David},
  journal={The Annals of Applied Probability}, volume = {21},
  pages={1663--1693},
  year={2011},
  publisher={JSTOR}
}

@article{ettinger2014,
 ISSN = {10505164},
 URL = {http://www.jstor.org/stable/42919723},
 author = {Boris Ettinger and Steven N. Evans and Alexandru Hening},
 journal = {The Annals of Applied Probability},
 number = {1},
 pages = {1--33},
 publisher = {Institute of Mathematical Statistics},
 title = {KILLED {B}ROWNIAN MOTION WITH A PRESCRIBED LIFETIME DISTRIBUTION AND MODELS OF DEFAULT},
 volume = {24},
 year = {2014}
}

@article{abundo2025,
  title     = "Inverse first-passage problems of a diffusion with resetting",
  author    = "Abundo, Mario",
  journal   = "Theory Probab. Math. Statist.",
  publisher = "American Mathematical Society (AMS)",
  volume    =  112,
  number    =  0,
  pages     = "17--36",
  month     =  may,
  year      =  2025
}

@article{Montroll1965,
    author = {Montroll, Elliott W. and Weiss, George H.},
    title = {Random Walks on Lattices. II},
    journal = {Journal of Mathematical Physics},
    volume = {6},
    number = {2},
    pages = {167-181},
    year = {1965},
    month = {02},
    issn = {0022-2488},
    doi = {10.1063/1.1704269},
    url = {https://doi.org/10.1063/1.1704269}
}

@article{Metzler2000,
title = {The random walk's guide to anomalous diffusion: a fractional dynamics approach},
journal = {Physics Reports},
volume = {339},
number = {1},
pages = {1-77},
year = {2000},
issn = {0370-1573},
doi = {https://doi.org/10.1016/S0370-1573(00)00070-3},
url = {https://www.sciencedirect.com/science/article/pii/S0370157300000703},
author = {Ralf Metzler and Joseph Klafter}
}

@article{Gavrilov2017,
    author = {Gavrilov, Momčilo and Bechhoefer, John},
    title = {Feedback traps for virtual potentials},
    journal = {Philosophical Transactions of the Royal Society A: Mathematical, Physical and Engineering Sciences},
    volume = {375},
    number = {2088},
    pages = {20160217},
    year = {2017},
    month = {03},
    issn = {1364-503X},
    doi = {10.1098/rsta.2016.0217},
    url = {https://doi.org/10.1098/rsta.2016.0217},
}

@article{Cohen2005,
  title = {Control of Nanoparticles with Arbitrary Two-Dimensional Force Fields},
  author = {Cohen, Adam E.},
  journal = {Phys. Rev. Lett.},
  volume = {94},
  issue = {11},
  pages = {118102},
  numpages = {4},
  year = {2005},
  month = {Mar},
  publisher = {American Physical Society},
  doi = {10.1103/PhysRevLett.94.118102},
  url = {https://link.aps.org/doi/10.1103/PhysRevLett.94.118102}
}

@article{Grier2006,
author = {David G. Grier and Yael Roichman},
journal = {Appl. Opt.},
number = {5},
pages = {880--887},
publisher = {Optica Publishing Group},
title = {Holographic optical trapping},
volume = {45},
month = {Feb},
year = {2006},
url = {https://opg.optica.org/ao/abstract.cfm?URI=ao-45-5-880},
doi = {10.1364/AO.45.000880},
}

@article{Satija2020,
author = {Rohit Satija  and Alexander M. Berezhkovskii  and Dmitrii E. Makarov },
title = {Broad distributions of transition-path times are fingerprints of multidimensionality of the underlying free energy landscapes},
journal = {Proceedings of the National Academy of Sciences},
volume = {117},
number = {44},
pages = {27116-27123},
year = {2020},
doi = {10.1073/pnas.2008307117},
URL = {https://www.pnas.org/doi/abs/10.1073/pnas.2008307117}}

@article{Kolomeisky2011,
    author = {Kolomeisky, Anatoly B.},
    title = {Physics of protein–{DNA} interactions: mechanisms of facilitated target search},
    journal = {Physical Chemistry Chemical Physics},
    volume = {13},
    number = {6},
    pages = {2088-2095},
    year = {2011},
    month = {02},
    issn = {1463-9076},
    doi = {10.1039/c0cp01966f},
    url = {https://doi.org/10.1039/c0cp01966f}
}

@misc{SM,
  note = {See Supplemental Material at [URL will be inserted by publisher] for
          additional derivations. This includes: i) Deriving Eqs. (3) and (4) using the flux-over-population approach, ii) Solving the backward equation for the MFPT, and iii) The continuous-time random walk analysis of the example in Fig. 2.
          
          This material includes and additional reference, Ref.~\cite{Little1961}},
}
